# KEPLER RAPIDLY ROTATING GIANT STARS

A. D. Costa[1], B. L. Canto Martins[1], J. P. Bravo[1], F. Paz-Chinchón[1], M. L. das Chagas[1], I. C. Leao[1], G. Pereira de Oliveira[1], R. Rodrigues da Silva[1], S. Roque[1], L. L. A. de Oliveira [1], D. Freire da Silva [1], J. R. De Medeiros[1]

[1]Departamento de Física Teórica e Experimental, Universidade Federal do Rio Grande do Norte, Campus Universitário, Natal RN

renan@dfte.ufrn.br

## ABSTRACT

Rapidly rotating giant stars are relatively rare and may represent important stages of stellar evolution, resulting from stellar coalescence of close binary systems or accretion of sub-stellar companions by their hosting stars. In the present letter we report 17 giant stars observed in the scope of the Kepler space mission exhibiting rapid rotation behavior. For the first time the abnormal rotational behavior for this puzzling family of stars is revealed by direct measurements of rotation, namely from photometric rotation period, exhibiting very short rotation period with values ranging from 13 to 55 days. This finding points for remarkable surface rotation rates, up to 18 times the Sun rotation. These giants are combined with 6 other recently listed in the literature for mid-IR diagnostic based on WISE information, from which a trend for an infrared excess is revealed for at least a half of the stars, but at a level far lower than the dust excess emission shown by planet-bearing main-sequence stars.

## 1. INTRODUCTION

The literature reports a growing list of rapidly rotating giant stars that violate the general rule for slow rotation of normal giant stars predicted by stellar evolutionary theory (Rodrigues da Silva et al. 2015; Carlberg et al. 2011, 2012; Bidelman & MacConell 1973; Fekel et al. 1986; Fekel & Balachandran 1993; Fekel 1987; De Medeiros & Mayor 1999). Essentially, these stars present a single behavior, with constant radial velocity, and rotational velocity $v \sin i$ from moderate up to 50 times the solar rotation, in contrast to single giants with a normal rotating behavior, which shows $v \sin i$ values of a few $\mathrm{kms}^{-1}$. The yellow giant FK Comae, the King of Spin (Ayres et al. 2006), is the leading example of such an abnormal rotation with $P_{rot} = 2.412$ d (Chugainov 1966) and $v \sin i = 162.5\,\mathrm{kms}^{-1}$ (Huenemoerder et al. 1993). Following this stellar royalty, one can define the G5III-IV giant HD 199178 as the Spin Vice-king because of its $v \sin i = 80\,\mathrm{kms}^{-1}$ (Herbig 1958; Huenemoerder et al. 1993) and $P_{rot} = 3.337484$ d (Bopp et al. 1983).

In spite of a number of suggestions, recent stellar mergers (Bopp & Stencel 1981), sudden dredge–up of angular momentum from stellar interior (Simon & Drake 1989) and accretion of substellar companions (Peterson et al. 1983; Siess & Livio 1999a,b), the nature of such an abnormal rotation is not yet well understood. Carlberg et al. (2012) have claimed for consistent evidences that the accretion of giant planets of a few Jupiter masses could produce the rapid rotation of the referred rapid rotators, in agreement with theoretical predictions (e.g.: Siess & Livio 1999a,b), whereas Rodrigues da Silva et al. (2015) have reported that some of these abnormal giants exhibit an IRAS infrared (IR) excess, which, in principle, could be related to circumstellar dust. Rodrigues da Silva et al. (2015) have also reported the presence of these abnormal stars in classes of evolved stars other than the giants of luminosity class III, with the discovery of single rapid rotators among subgiant, bright giant and Ib supergiant stars.

As a rule, all the rapidly rotating single evolved stars reported to date, have had their rotational behavior defined from measurements of the projected rotational velocity v sin i. This study brings a pioneering analysis of the infrared behavior of a sample of 23 rapidly rotating giants, revealed by their very short photometric rotation period computed from Kepler light curves. This letter is organized as follows. Section 2 presents the Kepler data, with a brief description of the light curves treatment. Main results are presented in Sect. 3, with conclusions in Sect. 4.

## 2. STELLAR WORKING SAMPLE AND LIGHT CURVE ANALYSIS

The sample is composed by 23 stars observed in the scope of the Kepler mission (Koch et al. 2010), 17 of them first pointed out as potential rapidly rotating giants by Pinsonneault et al. (2014) and 6 by Tayar et al. (2015), from a total sample of 1916 giant stars classified from asteroseismic diagnostics. Indeed, the sample of rapid rotators from Tayar et al. (2015) is composed by 10 stars, but with 3 clear eclipsing binaries (KIC 3955867, KIC 4473933 and KIC 5193386). The final list of stars is given in Table 1 and displayed in Fig. 1 (top panel); their locations in the HR diagram are compatible with the red giant branch stage.

The stars were first analyzed using the Pre-Search Data Conditioning (PDC) light curves (LCs) from Kepler, which has been treated with the standard pipeline to remove thermal and kinematic trends (Jenkins et al. 2010). However some trends persist and must be manually removed for the sake of a better identification of the stellar LC variability (Paz et al. 2015). In total, 58 regions were removed from the sample. After this procedure, the Lomb-Scargle periodogram (Lomb 1976, Scargle 1982) was computed for each star and main periods with corresponding peaks having significance level greater than 99% being considered. Prior to assuming a period as valid, a careful visual inspection was performed based on the criteria by De Medeiros et al. (2013), from which we identified 9 stars having a periodic semi-sinusoidal behavior and sufficiently well behaved phase plots. For these stars we computed unambiguous rotation periods, with a signal-to-noise ratio greater than 0.001, and comprising only primary and secondary periodogram peaks, except for KIC 7108646 for which the third periodogram peak was selected as the variability peak. Here the "signal" is set as the amplitude of fitted period, while the "noise" is the standard deviation of the difference between nearest-neighbor LC measurements (see § 2 in De Medeiros et al. 2013). The computed periodograms are displayed in the right panels of Fig. 2 (KIC 2305930, KIC 4348593 and KIC 4937056), Fig. 3 (KIC 2305930, KIC 4348593 and KIC 4937056) and Fig. 4 (KIC 7103951, KIC 7108646 and KIC 8085964). Following the same recipe, we computed the periods for the 6 stars from Tayar et al. (2015) obtaining values very close to those given by the referred authors.

For 8 stars with no period values in Table 1 (KIC 3216467, KIC 3937217, KIC 4937011, KIC 6425652, KIC 6430534, KIC 7431665, KIC 7670419, KIC 9881628), in spite of the presence of apparent features in their LCs pointing for short period modulation it was not possible to extract reliable variability periods. Typically, these LCs present very low amplitudes in relation to their noise levels, thus showing faint peaks in their periodograms that cannot be distinguished from artifacts.

We also analyzed the periodicity in the 15 stars of present sample, with unambiguous period, using the wavelet technique (Grossman & Morlet 1984), applying the same procedure described in Bravo et al. (2014), where the 6th-order Morlet wavelet is used. The global spectrum of the signal was computed by averaging the local spectrum over time. To prevent some contribution from data gaps in the wavelet map energy distribution as misrepresenting relevant periodicities, the wavelet analysis is performed separately for each continuous data range for targets KIC 4348593, KIC 4937056, KIC 7103951, KIC 7108646,

KIC 8085964 and KIC 9469165. The global spectrum in these cases is then obtained adding the weighted average by time span of these parts. Typical wavelet local and global spectra for 9 stars are displayed in Figs. 2, 3, and 4, with the periods given in Table 1. The rotation periods $P_{rot}^{W}$ obtained by this analysis, which show some stability over time in wavelet maps, are in accordance with those obtained from Lomb-Scargle periodograms ($P_{rot}^{F}$ ). Although, for KIC 4937056 and 8825444, two possible rotation periods are considered since each one is persistent in different time intervals. Even though, these periodicities confirm that these stars may be classified as rapid rotators.

Finally we have computed the rotational velocity $v \sin i$ for the 9 stars from Pinsonneault et al. (2014), using computed rotation periods and radii (Fekel et al. 1986). The obtained values confirm that these stars follow the criteria used by different authors to define what constitutes rapid rotators (Rodrigues da Silva et al. 2015; Carlsberg et al. 2011; Fekel 1997), namely $v \sin i \geq 10\,\mathrm{kms}^{-1}$. This is the cutoff applied by Tayar et al. (2015) to define their rapid rotators, also analyzed in the present study.

## 3. RESULTS

To date, the rapidly rotating evolved stars have had their rotation behavior defined from $v \sin i$ projected rotational velocity measurements. The present study reveals this abnormal rotation behavior for a sample of 17 Kepler giant stars, on the basis of direct rotation measurements. These stars were first identified as rapidly rotating giants by Pinsonneault et al. (2014) on the basis of asteroseismic diagnostics. Here we confirm such a behavior by computing the rotation period on the basis of two independent techniques, Lomb-Scargle periodograms and wavelet procedures. Combining this sample with 6 Kepler giants defined by Tayar et al. (2015) as rapid rotators, we show a period range from 13 to 55 days, corresponding to a rotational velocity $v \sin i$ spread from 28 to 10 $\mathrm{kms}^{-1}$. These stars have masses ranging from 0.83 to 3.23 M and radius into the range 9.5-13.1 R , both parameters computed from asteroseismic diagnostics (Pinsonneault et al. 2014). The computed periods and radii points for striking rapid surface rotations, ranging from about 6 to 18 times the solar surface rotation.

### 3.1. ON THE MID-IR BEHAVIOR OF THE KEPLER RAPIDLY ROTATING GIANT STARS

As underlined, the nature of the abnormal rapidly rotating giants is essentially unknown to date, whilst some mechanisms have been proposed, including recent stellar mergers, sudden dredge-up of angular momentum from stellar interior, and the accretion of substellar companions. An interesting aspect to be underlined is that all these scenarios may produce circumstellar dust (e.g.: Siess and Livio 1999a,b; Zuckerman 2001).

The WISE, Wide-field Infrared Survey Explorer all-sky data (Wright et al. 2010), with observations centered at 3.4, 4.6, 12, and 22μm wavelengths, offers an unique opportunity for the study of the mid–IR behavior for a plethora of astronomical bodies, including different stellar families. Indeed, Lawler & Gladman (2012) studied the IR behavior for a sample of more than 900 Kepler targets, composed by stars with confirmed planets and stars with potential planet candidates, on the basis of WISE and 2MASS data, from where it was possible to identify 8 stars with mid-IR excess. Morales et al. (2012) conducted a similar study for a sample of 591 stars hosting planets from the Extrasolar Exoplanet Encyclopaedia (Schneider et al. 2011), with 9 stars presenting a mid-IR emission.

In order to study the IR characteristics of the present sample of rapid rotators in more details, we combined IR information from WISE and 2MASS data basis, following the approach by Wu et al. (2013) in the search for mid-IR excess. Fig. 1 (bottom panel) displays the color-color diagram, 2MASS $\mathbf{J} - \mathrm{H}$ versus WISE $\mathrm{K} - [22]$ for our working sample of 23 Kepler giants (blue circles and squares), irrespective of the quality of rotation period. According to the criteria applied by Wu et al. (2013), stars with mid-IR excess present a $\mathrm{K} - [22]$ larger than 0.22 for a $\mathbf{J} - \mathrm{H} > 0.3$, $\mathrm{K} - [22]$ larger than 0.21 for $0.1 < \mathbf{J} - \mathrm{H} \leq 0.3$ and $\mathrm{K} - [22]$ larger than 0.26 for $\mathbf{J} - \mathrm{H} \leq 0.1$ (dashed vertical red line). On the basis of these criteria at least a half of the stars in our sample of Kepler giants present a tendency for an emission excess at 22μm. For comparison, Fig. 1 (bottom panel) shows also the 8 stars from Lawler and Gladman (2012) (red circles) and the 7 stars from Morales et al. (2012) (green circles) with mid-IR emission excess, as well as 58 subgiant (black filled circles) and 40 giant (black open circles) stars with planet, listed in the Extrasolar Exoplanet Encyclopaedia (Schneider et al. 2011), with WISE and 2MASS photometry available.

The IR behavior of the Kepler rapid rotators in the color-color diagram 2MASS $\mathbf{J} - \mathrm{H}$ versus WISE $\mathrm{K} - [22]$, with regard to the other family of stars also displayed in Fig. 1 (bottom panel), points for an interesting scenario. These rapidly rotating giants present two different features with a group of stars showing a trend for an IR excess, confirming the results by Rodrigues da Silva et al. (2015), and another group showing a standard behavior with no IR excess. Nevertheless, the IR excess among the rapidly rotating giants is largely weaker than that observed in main-sequence stars hosting planets. For these latter, the IR excess is compatible with presence of warm to hot dust (e.g.: Lawler & Gladman 2012; Morales et al. 2012). Also interesting, is the group of rapidly rotating giants without IR excess, which follows the same behavior of the subgiant and giant stars hosting planets in agreement with a simple blackbody emission.

A particular remark is required for 7 subgiants with planets (CoRoT-21, CoRoT-22, CoRoT-26, Kepler-39, Kepler-40, Kepler-43 and Kepler-44): Their apparent IR excess should be taken with cautious because their WISE data presents low S/N values and a visual inspection of WISE images of the referred stars reveals that their W4 data are contaminated by background IR emission. For HD 199178 the apparent excess is just a diffraction artifact.

## 4. CONCLUSIONS

The discovery of G– and K–type apparently single evolved stars rotating fast in excess in relation of the theoretical predictions, represents an exciting topic for study in stellar astronomy. In this letter we report 23 Kepler giant stars presenting such an abnormal behavior, using, for the first time, rotation period measurements for the diagnostic of the rotational characteristics for 17 of them. This represents 1.2% of all giant stars listed by Pinsonneault et al. (2014). The computed rotation periods range from about 13 to 55 days, corresponding, respectively to rotational velocities of about 28 to 10 $kms^{-1}$, pointing for striking rapid surface rotations, up to 18 times the solar rotation. Clearly, these remarkable rotation rates are strongly faster than the values predicted by stellar evolutionary theory.

By combining WISE and 2MASS Infrared information, we shown that a fraction of the reported Kepler rapidly rotating giants present a tendency for a mid–IR excess. One interesting aspect in the present analysis emerges from the comparison between the corresponding IR behavior for these Kepler giants and main–sequence planet host stars showing dust excess emission, on the basis of the same IR excess diagnostic. The presence of two IR excess distinct levels is clearly observed for the rapidly rotating giants and stars with planets, with the mid–IR excess level for the stars with planets largely higher than that observed in the rapidly rotating giant. Nonetheless, the present study shows that mid–IR excess is not a general rule among rapidly rotating giants, with a fraction of the analyzed Kepler rapid rotators showing a standard behavior, without IR excess, in agreement with a simple blackbody emission. As it is now well established, a mid-IR excess points, in general, for the existence of warm circumstellar dust, an aspect that could also explain the observed IR excess for some Kepler rapidly rotating giant here reported. However the lack of a mid–IR excess does not necessarily reflet the absence of circumstellar material, because cool discs are detectable only at longer wavelengths, with very little or no excess in the mid-IR.

Research activities of the Stellar Board of the Federal University of Rio Grande do Norte are supported by CNPq and FAPERN brazilian agencies. MLC, JPB and RRS acknowledge Post-doctorate fellowships of the CAPES brazilian agency. ICL acknowledges CNPq and CAPES PNPD fellowships. ADC, FPC, GPO, SR, LLAO and DFS acknowledge graduate fellowships of CAPES. We also acknowledge the INCT INEspaço for partial financial support. The authors warmly thank the Kepler and WISE Technical and Management Staff for the development, operation, maintenance and success of the Missions.

---

This manuscript was prepared with the AAS LaTeX macros v5.2.

Table 1. Stellar parameters, including colors $J-H$ (from 2MASS) and $K-[22]$ (from WISE) and rotation periods computed via Fourier $\left(P^F_{rot}\right)$ and Wavelet $\left(P^W_{rot}\right)$.

| Star | $T_{eff}$ (K) | Mass ($M_\odot$) | Radius ($R_\odot$) | $v \sin i$ ($km s^{-1}$) | $\log g$ ($cm/s^2$) | $J-H$ (dex) | $K-[22]$ (dex) | $P^F_{rot}$ (days) | $P^W_{rot}$ (days) |
|---|---|---|---|---|---|---|---|---|---|
| Pinsonneault et al. (2014) | | | | | | | | | |
| KIC2305930 | 4716 ± 107 | 0.87 | 10.16 | 12.19 | $2.367^{+0.014}_{-0.011}$ | 0.528 | 0.731 | 33.12 | 32.62 |
| KIC3216467 | 4389 ± 70 | 1.11 | 11.26 | ... | $2.382^{+0.014}_{-0.014}$ | 0.612 | 0.047 | ... | ... |
| KIC3937217 | 4616 ± 93 | 1.01 | 11.02 | ... | $2.358^{+0.013}_{-0.013}$ | 0.557 | −0.038 | ... | ... |
| KIC4348593 | 5018 ± 82 | 3.23 | 13.12 | 14.41 | $2.707^{+0.017}_{-0.018}$ | 0.440 | 0.075 | 36.17 | 35.02 |
| KIC4937011 | 4558 ± 86 | ... | 9.07[a] | ... | 2.680[a] | 0.549 | 1.317 | ... | ... |
| KIC4937056 | 4750 ± 83 | 1.65 | 11.02 | 16.01 | $2.572^{+0.014}_{-0.018}$ | 0.513 | 0.846 | 27.36 | 17.21 / 26.08 |
| KIC6425652 | 4620 ± 95 | 0.88 | 10.56 | ... | $2.343^{+0.011}_{-0.010}$ | 0.569 | 1.375 | ... | ... |
| KIC6430534 | 4444 ± 81 | 1.01 | 11.29 | ... | $2.339^{+0.013}_{-0.011}$ | 0.592 | 0.270 | ... | ... |
| KIC7103951 | 4801 ± 85 | ... | 11.56[a] | 12.53 | 2.502[a] | 0.408 | 0.065 | 36.65 | 37.45 |
| KIC7108646 | 4513 ± 87 | 1.06 | 11.08 | 22.40 | $2.372^{+0.025}_{-0.020}$ | 0.599 | 0.467 | 19.66 | 18.81 |
| KIC7431665 | 4548 ± 88 | ... | 9.02[a] | ... | 2.686[a] | 0.522 | 0.258 | ... | ... |
| KIC7670419 | 4444 ± 77 | 1.03 | 10.96 | ... | $2.373^{+0.019}_{-0.015}$ | 0.619 | 0.014 | ... | ... |
| KIC8085964 | 4945 ± 76 | 2.69 | 9.77 | 19.96 | $2.889^{+0.016}_{-0.017}$ | 0.415 | 0.045 | 19.45 | 20.63 |
| KIC8825444 | 4964 ± 79 | 2.75 | 10.58 | 14.20 | $2.833^{+0.013}_{-0.014}$ | 0.471 | 0.993 | 29.62 | 28.12 / 42.62 |
| KIC9469165 | 4806 ± 110 | 0.83 | 9.85 | 28.11 | $2.375^{+0.013}_{-0.013}$ | 0.506 | −0.119 | 13.93 | 13.94 |
| KIC9518802 | 4961 ± 87 | 2.44 | 9.48 | 10.19 | $2.870^{+0.013}_{-0.014}$ | 0.422 | 0.128 | 36.95 | 43.07 / 20.09 |
| KIC9881628 | 4724 ± 99 | 0.86 | 10.21 | ... | $2.361^{+0.011}_{-0.010}$ | 0.509 | −0.117 | ... | ... |
| Tayar et al. (2015) | | | | | | | | | |
| KIC2285032 | 4284 | ... | 14.43[a] | 21.63 | 2.72 | 0.649 | 0.347 | 39.89 | 40.16 |
| KIC10293335 | 4363 | 2.07 | 21.29[a] | 12.66 | 2.45 | 0.710 | −0.289 | 55.96 | 47.20 |
| KIC10417308 | 4734 | 1.09 | 8.38[a] | 11.22 | 2.96 | 0.506 | 0.432 | 40.15 | 38.57 |
| KIC11497421 | 4685 | 1.10 | 7.91[a] | 10.19 | 2.98 | 0.554 | 0.872 | 40.59 | 39.42 |
| KIC11597759 | 4658 | 0.91 | 6.36[a] | 11.24 | 2.98 | 0.537 | 0.648 | 54.74 | 47.16 |
| KIC12003253 | 4674 | 1.16 | 9.54[a] | 10.30 | 3.04 | 0.555 | 0.042 | 50.43 | 51.00 |

[a]Kepler Input Catalog (Kepler Mission Team, 2009)

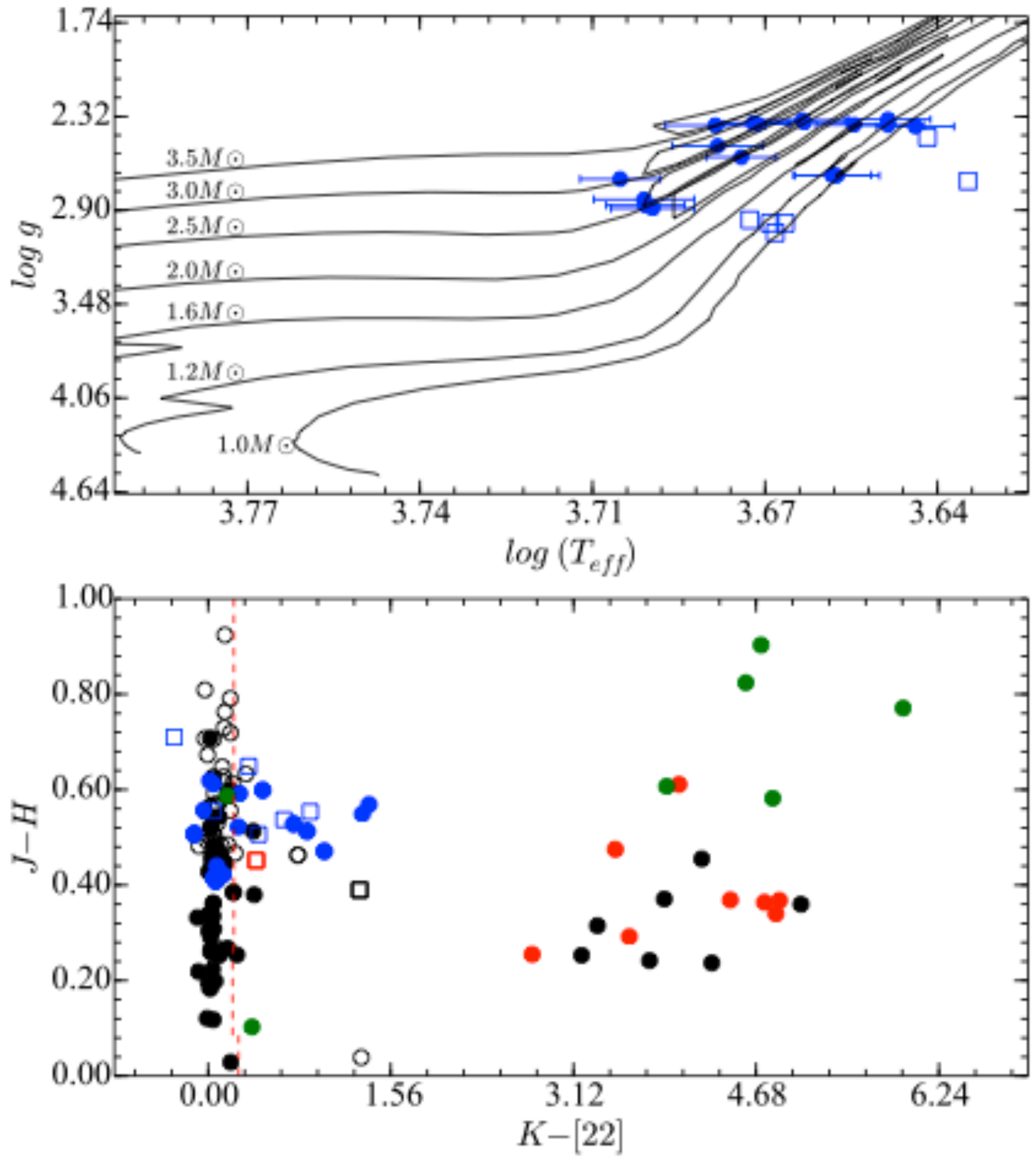

Fig. 1.— Top: Distribution of the 23 Kepler rapidly rotating giants in the HR diagram. Evolutionary tracks at $[Fe/H] = 0$, are from Girardi et al. (2000). Circles and open squares refer to stars from Pinsonneault et al. (2014) and Tayar et al. (2015), respectively. Bottom: Color-color diagram $J - H$ versus $K - [22]$ for the referred 23 giants (symbols as in the top panel), main-sequence stars with planets from Morales et al. (2012) (green circles) and from Lawler and Gladman (2012) (red circles), subgiants (black filled circles) and giants (black open circles) stars with planets from the Extrasolar Exoplanets Encyclopaedia. FK Comae and HD 199178 are indicated by red and black open squares, respectively. The red dashed line gives the criterion to define WISE 22µm excess from Wu et al. (2013), according which stellar excess candidates should populate the region with $K - [22]$ larger than about 0.2.

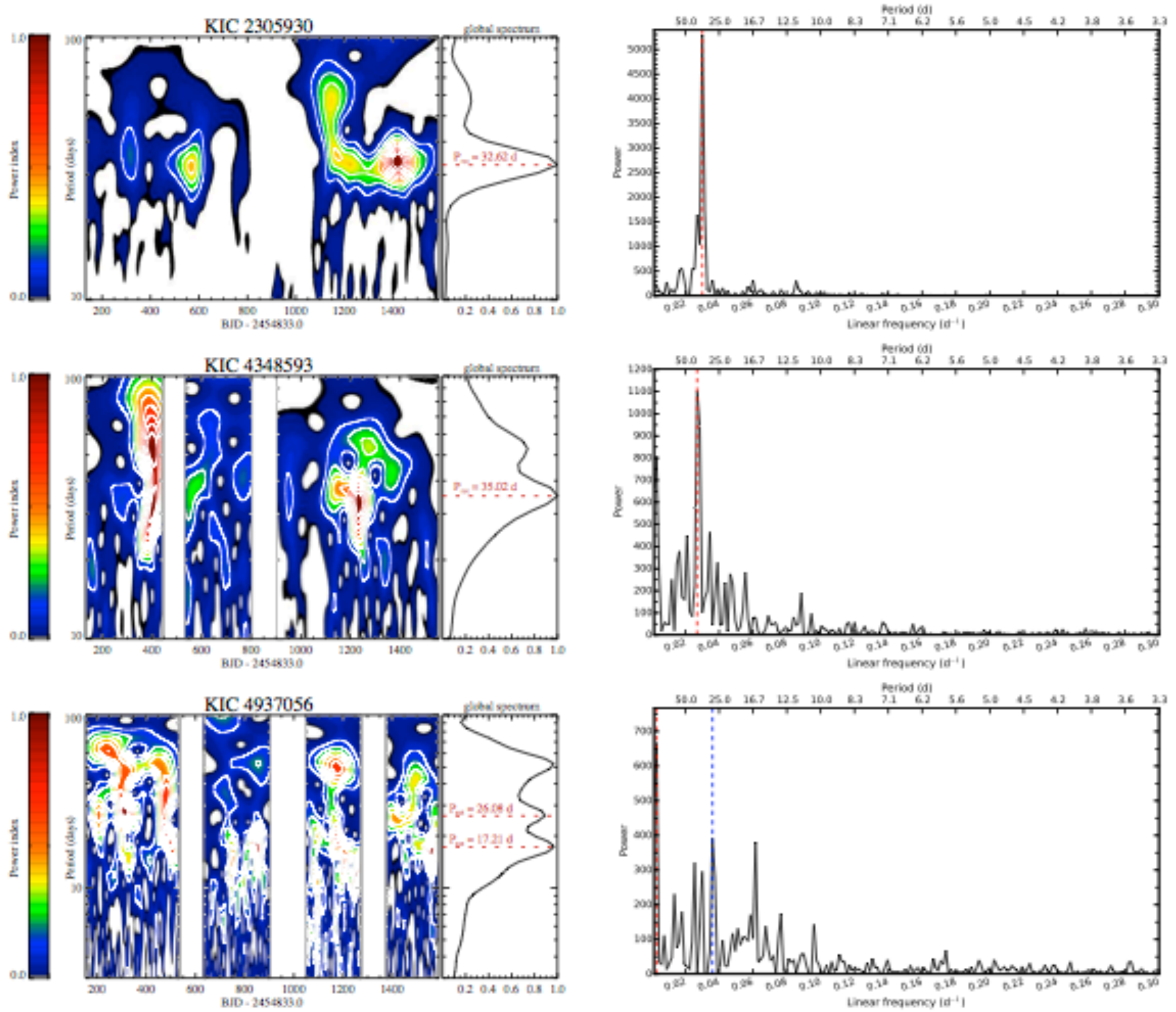

Fig. 2.— Left: Wavelet local and global spectra of Kepler giant stars (from top to bottom) KIC 2305930, KIC 4348593 and KIC 4937056. The rotation periods are illustrated by a red dashed line in the global spectrum. For KIC 4937056, two possible rotation periods are considered since each one is persistent in different time intervals. Contour levels are 90%, 80%, 70%,..., 20% and 10% of the map maximum. Right: Respective Lomb-Scargle periodograms for the same giant stars. The selected period and the periods having a higher power were marked with a vertical dashed line (in some cases the higher peak did not reflect a physical variability) as follows: first, second, and third peaks are marked with red, blue, and green, respectively. The x-axis represents linear frequency, while the auxiliary upper x-axis represents period. The y-axis represents the power of the periodogram.

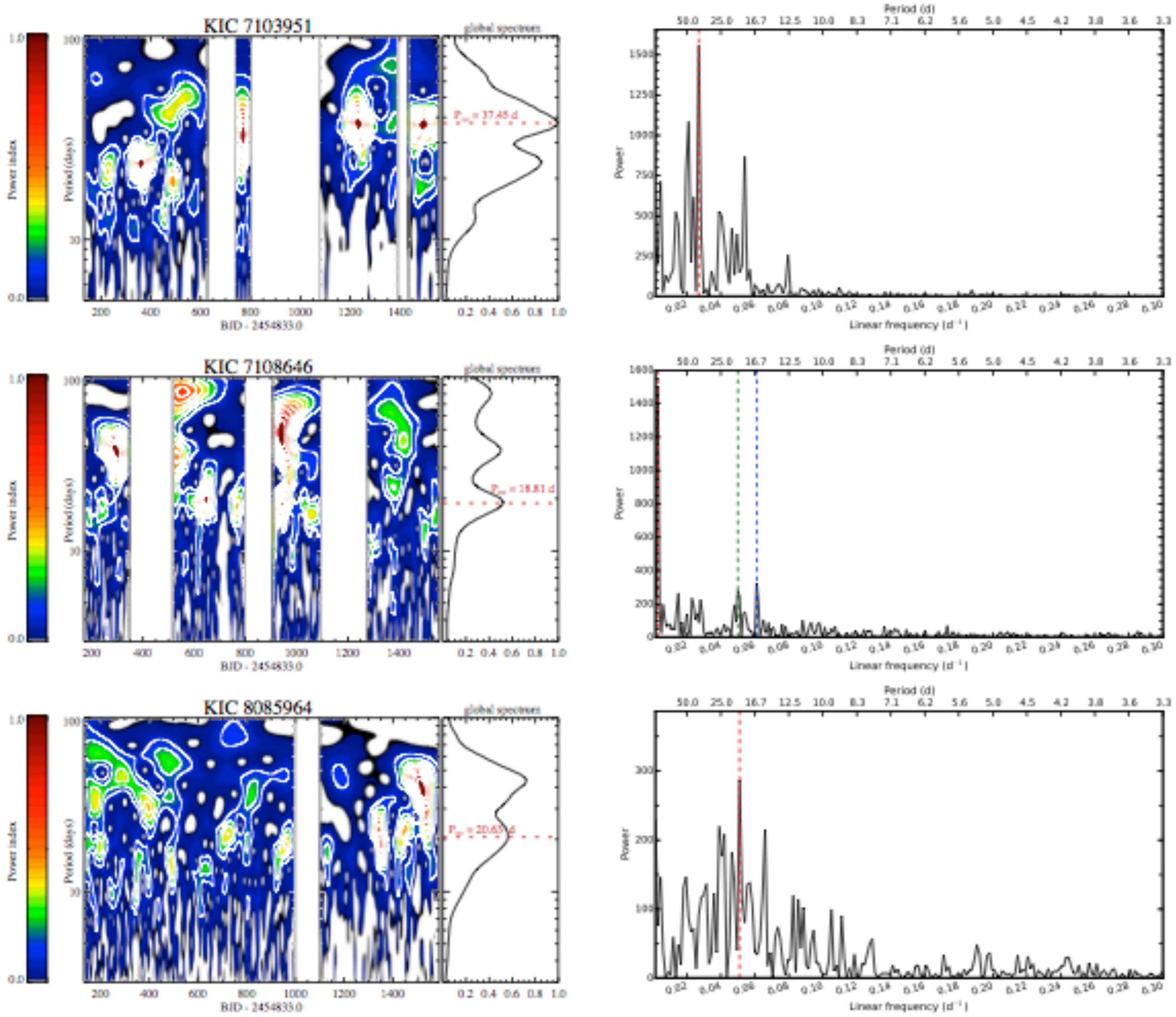

Fig. 3.— Same as Fig. 1 but for Kepler giants KIC 7103951, KIC 7108646 and KIC 8085964 (from top to bottom).

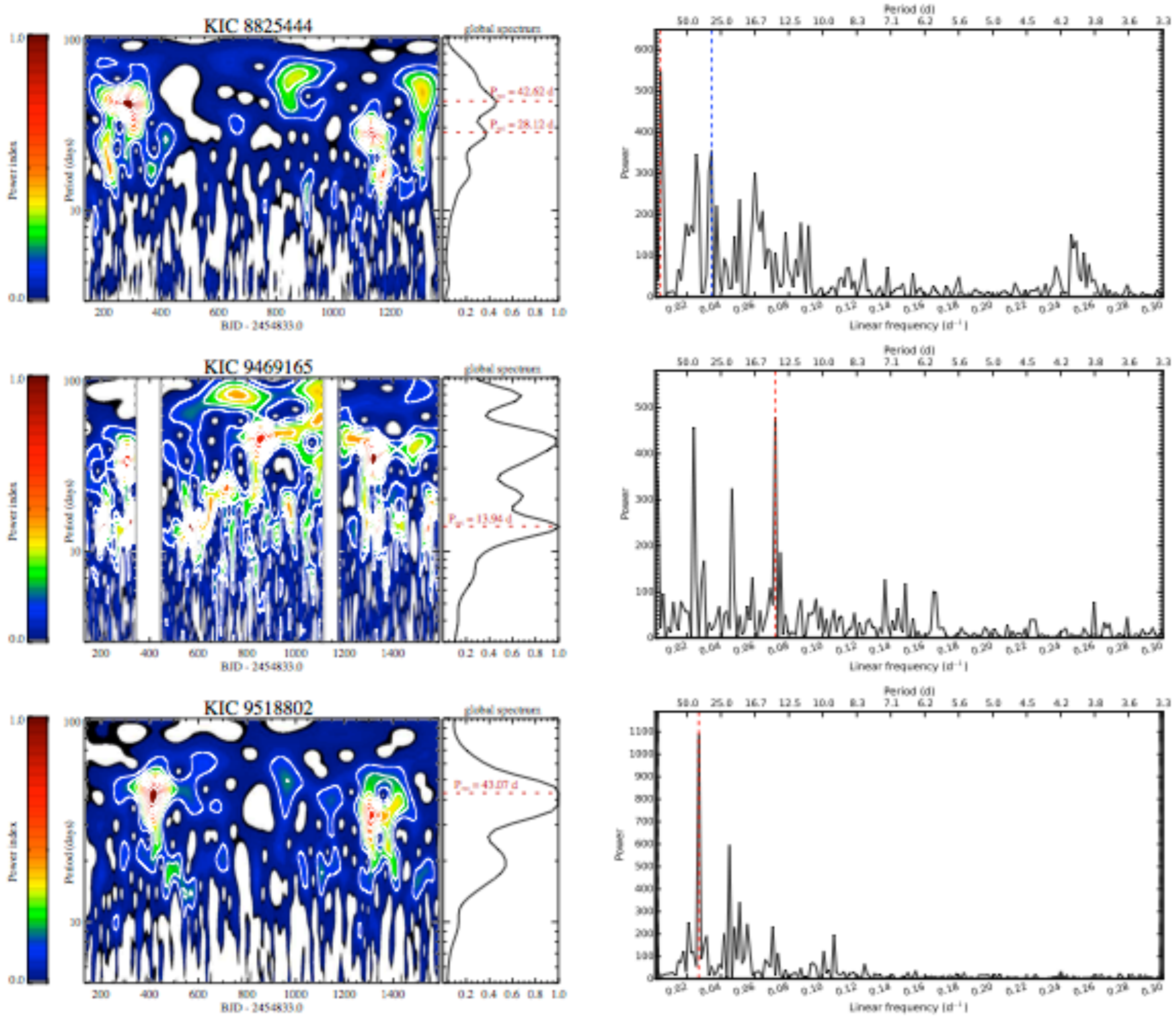

Fig. 4.— Same as Fig. 1 but for Kepler giants KIC 8825444, KIC 9469165 and KIC 9518802 (from top to bottom). For KIC 8825444, two possible periods are considered in the global spectrum since each one is persistent in different time intervals.